\documentclass[preprint,aps,showpacs,nofootinbib]{revtex4-1}
\usepackage{amsmath}
\usepackage{amsfonts}
\usepackage{amssymb,amscd,amsbsy}
\usepackage{epsfig}
\usepackage{color}

\usepackage{mathtools}

\newcommand{\bea}{\begin{eqnarray}}
\newcommand{\eea}{\end{eqnarray}}

\newcommand{\vect}[1]{\mathbf{#1}}

\newcommand{\kt}{k_{\rm B}T}

\newcommand{\cref}{c^{(2)}_{\rm ref}}

\newcommand{\nv}{n_{\rm vac}}

\begin{document}
\title{Stable and metastable hard sphere crystals in Fundamental Measure Theory}

\author{M. H. Yamani$^{1,2}$ and M. Oettel$^{2}$}
\affiliation{ $^1$ Johannes Gutenberg--Universit\"at Mainz, Institut f{\"ur} Physik,
  WA 331, D--55099 Mainz, Germany \\
$^2$ Institut f\"ur Angewandte Physik, Eberhard Karls--Universit\"at T\"ubingen, D--72076 T\"ubingen, Germany
}


\begin{abstract}
Using fully minimized fundamental measure functionals, we investigate free energies, vacancy concentrations and density
distributions for bcc, fcc and hcp hard--sphere crystals. Results are complemented by an approach
due to Stillinger which is based on expanding
the crystal partition function in terms of the number $n$ of free particles while the remaining particles
are frozen at their ideal lattice positions. The free energies of fcc/hcp and one branch of bcc agree well
with Stillinger's approach truncated at $n=2$. A second branch of bcc solutions features rather spread--out
density distributions around lattice sites and large equilibrium vacancy concentrations and is presumably 
linked to the shear instability of the bcc phase. Within fundamental measure theory and the Stillinger approach
($n=2$), hcp is more stable than fcc by a free energy per particle of about 0.001 $\kt$. In previous
simulation work, the reverse situation has been found which can be rationalized in terms of effects due to
a correlated motion of at least 5 particles in the Stillinger picture.      
\end{abstract}


\maketitle

\section{Introduction}

The crystal lattices of monatomic substances are very often of face--centered cubic (fcc), hexagonally close--packed (hcp)
or body--centered cubic (bcc) type. Still, it is a formidable problem in statistical 
mechanics and quantum chemistry to predict the stable crystal structure and its free energy for a given substance.
Approximating the particle interactions in this substance by classical two--body potentials makes the problem
amenable to a treatment using methods of classical statistical mechanics, most notably Monte Carlo (MC) simulations
and (classical) density functional theory (DFT). While the approximation using two--body potentials may not be very
accurate for truly atomic substances, the advance in colloid synthesis allows to realize systems with simple
two--body potentials to a good degree of approximation, thus colloid suspensions are a perfect model system for investigating
freezing in classical statistical mechanics.     

For isotropic two--body potentials $u(r)$ ($r$ is the center distance between two particles) 
a substantial amount of knowledge has been gathered. For potentials with a repulsive core
the steepness of the core mainly determines the stability of fcc over bcc, with fcc being more stable for steeper cores. This has
been investigated for power--law potentials $u \propto (1/r)^n$ \cite{Dav05} and screened exponentials $u \propto \exp(-\kappa r)/r$ \cite{Mei91,Hei13}
where the parameters $n,\kappa$ determine the steepness of the potential. In the hard--sphere limit ($n,\kappa \to \infty$), fcc
appears to be the stable, equilibrium structure and a possible bcc structure is unstable against small shear \cite{Hoo72} which
is reflected in squared phonon frequencies $\omega^2(\vect k)$ being negative for certain wave vectors $\vect k$.   

For hard spheres, it is a much more delicate issue whether fcc is more stable than other close--packing structures, most
notably hcp. Early theoretical work by Stillinger {\em et al.} analyzed the free energy of hard disks and fcc and hcp hard sphere crystals in terms of an
expansion in the number $n$ of contiguous particles (free to move) in an otherwise frozen matrix of particles at their
ideal lattice positions \cite{Sti65,Sti67,Sti68} (see below). This expansion could be done analytically only for densities in the vicinity of close--packing 
and, for $n=2$ and $n=3$ (by quite a {\em tour de force}), resulted in hcp being more stable than fcc
by a free energy difference per particle $\Delta F/N \sim 10^{-3}$ $\kt$. However, the individual terms contributing in this series
are much larger than this value of $\Delta F/N$. An extension of this method \cite{Koch05} (still only near close--packing)
to $n=5$ shows the reverse situation: fcc is more stable than hcp and  $\Delta F/N \sim -10^{-3}$ $\kt$, but the
last term in the series is still larger in magnitude than  $\Delta F/N$ (about 6 times for fcc and 3 times for hcp).
Simulation work confirms the stability of fcc over hcp also for smaller densities (around coexistence). Using
a single--occupancy cell (SOC) method, Ref.~\cite{Woo97} estimates $\Delta F/N = -(5 \pm 1)\cdot 10^{-3}$ $\kt$ at
a density of $\rho_0\sigma^3=1.041$ (approximately at coexistence, $\sigma$ is the hard sphere diameter).  
In this method, particles are constrained to their Wigner--Seitz cells and the free energy difference is found by
integrating the equation of state. The limitations of this method could be overcome by the powerful Monte--Carlo (MC)
lattice switch method which allows to compute directly the free energy difference between two different lattice structures
\cite{Wil97}. At $\rho_0\sigma^3=1.10$ the result is $\Delta F/N  = -(0.86 \pm 0.03) \cdot 10^{-3}$ $\kt$.  
Thus the result of the high--density Stillinger series for $n=5$ for the stability of fcc over hcp and the magnitude of the free energy difference 
is consistent with the MC simulation result at a considerably smaller density. One may tentatively conclude that for all
densities the stability of
fcc in the hard sphere system is a subtle result of the correlated movement of five and more particles and the effect in the free energy
is very small.   

In view of this evidence it appears to be very hard to contribute to the theoretical understanding of the stability of
fcc over hcp beyond the Stillinger arguments. In this respect, density functional theory (DFT) seems to be the only promising
candidate theory.  In the general framework of classical DFT
crystals are viewed as ``self--sustained'', periodic density oscillations of a liquid, which minimize
a unique, but in general unknown free energy functional. Ramakrishnan and Yussouff demonstrated \cite{Ram79} that
a simple functional, which is Taylor--expanded about a homogeneous liquid state near coexistence semi--quantitatively
accounts for the freezing transition in the hard sphere system. Such Taylor--expanded functionals can be
devised for a wide range of two--particle potentials but they are often not very precise. Nevertheless they are
a useful starting point for deriving more coarse--grained models via gradient expansions leading to
phase field crystal models for materials science \cite{Low12}. For hard--body potentials
there is a constructive way to derive functionals ``from scratch'' (not relying on
perturbative expansions) using essentially
geometric arguments. This approach is known as fundamental measure theory (FMT) \cite{Ros89,Han06,Kor12}. With regard to the
description of crystals, it has proved  to be fruitful to consider the  zero--dimensional (0D) limit
of density distributions localized to a point and their exactly known free energy \cite{Tar97}.
By requiring that the density functional reproduces this 0D free energy for density peaks at one, two
and three points in space, a density functional may be constructed which exhibits solid phase properties in
very good agreement with simulations \cite{Tar00}. (In the case of density distributions with $\delta$--peaks at
three points, the 0D free energy is reproduced only approximately.)   

In the seminal work \cite{Tar00}, the crystal density distributions were parametrized with isotropic Gaussians with
variable width parameter and normalization (to allow for a finite vacancy concentration $\nv$). By minimizing the 
free energy with respect to the width parameter and the normalization, the following results were obtained:
The crystal free energy per particle $F/N$ agrees with simulation to within less than a percent and the Gaussian
width is only slightly smaller than seen in simulations. However, furthermore it was found: 
Z(i) No free energy minimum for $\nv>0$ and (ii) equal free energies for fcc and hcp.  
In a study combining simulation and free minimization of FMT functionals \cite{Oet10} it was shown that
(i) is a defect of the functional used in \cite{Tar00} (the Tarazona tensor functional) and that
upon free minimization the White Bear II tensor functional of Ref.~\cite{Han06} gives thermodynamically
consistent results\footnote{This is discussed in Ref.~\cite{Oet10}, Sec. III.A. under the heading ``$\mu$ consistency''.}  
with a small equilibrium vacancy concentration $\nv \sim 2 \cdot 10^{-5}$ for fcc.
The free energies per particle obtained by free minimization vs. constrained minimization
using isotropic Gaussians differ by about $2\cdot 10^{-3}$ $\kt$ (near coexistence), which is of the order 
of magnitude one would also expect for the fcc--hcp difference $\Delta F/N$. Hence one is lead to the
suspicion that (ii) (i.e. $\Delta F/N=0$) is an artefact of the constrained minimization. This issue will be addressed here. 

Apart from the issue of fcc vs. hcp in hard spheres, FMT is also suited to investigate the metastable bcc crystal
(which in FMT is simply stabilized by the periodic boundary conditions). A previous FMT study \cite{Lut06} using constrained minimization
found two metastable bcc branches as well as a peculiar behavior of the lattice site density peaks when the density is increased.
We will investigate this finding further by fully minimizing the FMT functional and will relate our results 
to the Stillinger series.

The article will be structured as follows: We recapitulate basic FMT as used here (Sec.~\ref{sec:fmt}) and 
Stillinger's expansion in correlated, contiguous particles (Sec.~\ref{sec:still}). Results from both approaches 
are presented in Sec.~\ref{sec:results} and Sec.~\ref{sec:summary}  summarizes and concludes our work.



\section{Theory}

\subsection{Fundamental measure theory}
\label{sec:fmt}

\subsubsection{Definition of functionals} 

In the framework of density functional theory, the grand canonical free energy is a functional 
of the one-body density profile $\rho(\mathbf{r})$
\bea
 \Omega[\rho] = \mathcal{F}^{\rm id}[\rho] + \mathcal{F}^{\rm ex}[\rho]-\int d\mathbf{r} (\mu-V^{\rm ext}(\mathbf{r}))\rho(\mathbf{r})\;.
\eea
where $\mathcal{F}^{\rm id}$ and $\mathcal{F}^{\rm ex}$ denote the ideal and excess free energy 
functionals of the fluid. $\mu$ denotes the chemical potential 
and the external potential is represented by $V^{\rm ext}$. The exact form of the ideal part 
of the free energy is given by
\bea
\beta \mathcal{F}^{\rm id}[\rho]=\int d^{3}r \beta f^{\rm id}(\mathbf{r})=
\int d^{3}r\rho(\mathbf{r}) ( \ln[\Lambda^{3} \rho(\mathbf{r})]-1 )\;.
\eea
Here, $\Lambda$ is the de-Broglie wavelength and $\beta = 1/(\kt)$.

Fundamental measure theory (FMT) currently is the most 
precise functional for the excess free energy part for the hard sphere fluid. The corresponding
excess free energy is given by
\bea
 \mathcal{F}^{\rm ex} &=& \int d\mathbf{r} f^{\rm ex}(\{\mathbf{n}[\rho(\mathbf{r})]\}))\;, \\
 \beta f^{\rm ex}(\{\mathbf{n}[\rho(\mathbf{r})]\})) &=& n_{0} \ln(1-n_{3}) + \varphi_{1}(n_{3})
\frac{n_{1} n_{2} - \mathbf{n}_{1}\cdot \mathbf{n}_{2}}{1-n_{3}} \nonumber\\
& &+ \varphi_{2}(n_{3}) \frac{3\;(-n_{2}\;\mathbf{n_{2}} \cdot \mathbf{n_{2}} + n_{2,i}n^{t}_{ij}n_{2,j}+n_{2}n^{t}_{ij}n^{t}_{ji} 
- n^{t}_{ij}n^{t}_{jk}n^{t}_{ki} )}{16 \pi (1-n_{3})^{2}}\;.
\eea
Here, $f^{\rm ex}$ is the excess free energy density which is a 
(local) function of a set of weighted densities $\{\mathbf{n}(\mathbf{r})\}=\{n_{0},n_{1},n_{2},n_{3}, 
\mathbf{n}_{1}, \mathbf{n}_{2}, n_{T} \}$ with four scalar, two vector and one tensorial weighted densities. 
These are related to the density profile $\rho(\mathbf{r})$ by the convolutions $n_{\alpha}(\mathbf{r}) = \int d\mathbf{r}^{\prime}
\,\rho(\mathbf{r}^{\prime})\,w^{\alpha}(\mathbf{r}-\mathbf{r}^{\prime})$. 
The weight functions are given by ($R=\sigma/2$ is the hard sphere radius):
\bea
w^{3}(\mathbf{r}) &=& \Theta(R - r)\;,\nonumber\\
w^{2}(\mathbf{r}) &=& \delta(R - r)\;,\nonumber\\
w^{1}(\mathbf{r}) &=& w^{2}(\mathbf{r})/(4\pi R)\;,\nonumber\\
w^{0}(\mathbf{r}) &=& w^{2}(\mathbf{r})/(4\pi R^{2})\;,\\
\mathbf{w}^{2}(\mathbf{r}) &=& \mathbf{r}/r\;\delta(R - r)\;,\nonumber\\
\mathbf{w}^{1}(\mathbf{r}) &=& \mathbf{w}^{2}/(4\pi R)\;,\nonumber\\
w^{t}_{ij} &=& r_{i}r_{j} / \mathbf{r}^{2}\;\delta(R-r)\;.\nonumber
\eea
By choosing 
\bea
\varphi_{1}=1\quad \text{and} \quad \varphi_{2}=1
\eea
we obtain Tarazona's tensor functional~\cite{Tar00} based on the original Rosenfeld functional \cite{Ros89}.
The choice
\bea
\varphi_{1}&=&1 \;, \nonumber \\
\varphi_{2}&=&1-\frac{-2n_{3}+3n^{2}_{3}-2(1-n_{3})^{2}\text{ln}(1-n_{3})}{3n^{2}_{3}}
\eea
corresponds to the tensor version of the White Bear I functional~\cite{Roth02}. Finally, with
\bea
\varphi_{1}&=&1+\frac{2n_{3}-n^{2}_{3}+2(1-n_{3})\text{ln}(1-n_{3})}{3n^{2}_{3}}\nonumber \;, \\
\varphi_{2}&=&1-\frac{2n_{3}-3n^{2}_{3}+2n^{3}_{3}+2(1-n_{3})^{2}\text{ln}(1-n_{3})}{3n^{2}_{3}}\;,
\eea
the tensor version of the white Bear II functional is recovered~\cite{Han06}. This functional is most consistent with respect 
to restrictions 
imposed by morphological thermodynamics~\cite{Koen04}. 

In density functional theory, the crystal is viewed as a self--sustained inhomogeneous fluid. Therefore, beside 
bulk and inhomogeneous fluids, it is possible to study properties of the hard--sphere crystal within the 
framework of FMT.
Using the variational principle, the equilibrium density profile  $\rho_{\rm eq}(\mathbf{r})$ 
is determined via minimizing the grand canonical free energy functional
which leads to the Euler-Lagrange equation:
\bea
\label{ele.eq}
\beta^{-1}\ln{\frac{\rho_{\rm eq}(\mathbf{r})}{\rho_{0}}} = -\frac{\delta \mathcal{F}^{\rm ex}[\rho(\mathbf{r})]}
{\delta \rho(\mathbf{r})}+\mu^{\rm ex}-V^{\rm ext}(\mathbf{r})\;.
\eea
For the equilibrium crystal, $V^{\rm ext}(\mathbf{r}) =0$ and $\rho_{\rm eq}(\mathbf{r})$ is lattice--periodic,
and $\rho_{0}$, the homogeneous density (bulk density), is fixed by the excess chemical potential
$\mu^{\rm ex}$.
Being computationally simpler than a free minimization of the density profile, 
crystal density profiles are often obtained by a constrained minimization 
of a model profile with only a few free parameters such as e.g. a Gaussian profile
\bea
 \rho_{\rm cr}(\mathbf{r})=\sum_{\textrm{lattice\;sites\;$i$}}(1-n_{\rm vac})\biggl(\frac{\alpha}{\pi}\biggr)^{3/2}
\exp{\biggl(-\alpha(\mathbf{r}-\mathbf{r}_{i})^{2}\biggr)}\;.
 \label{eq:gauss}
\eea
Here, the free parameters are the Gaussian peak width $\alpha$ and the vacancy concentration $n_{\rm vac}$.

\subsubsection{Choice of unit cells for the numerical solution of Euler-Lagrange equation} 

\label{sec:fmt_num}

\begin{figure}[hb]
  \centering
  \includegraphics[width=8cm]{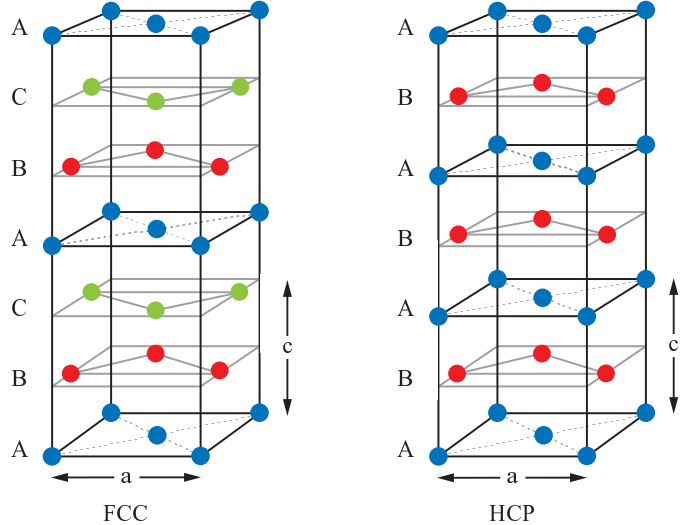}
 \caption{(color online) Extended unit cells for the fcc and hcp crystal structures. The fcc layers cycle among the three identical,
but laterally shifted layers, the blue A layer, the red B layer and the green C layer. 
For hcp, the A and B layers alternate. Positions of the lattice points of the first layers from the bottom are:\\
$\mathrm{layer\;A}:(0,0,0),\thinspace(a,0,0),\thinspace(0,\sqrt{3}\,a,0),\thinspace(a,\sqrt{3}\,a,0),\thinspace(\frac{1}{2}\,a,\frac{\sqrt{3}}{2}\,a,0),\;\; \mathrm{layer\;B}:(0,\frac{1}{\sqrt{3}}\,a,\frac{1}{2}\,c),\\ \thinspace(a,\frac{1}{\sqrt{3}}\,a,\frac{1}{2}\,c),\thinspace(\frac{1}{2}\,a,\frac{5}{2\sqrt{3}}\,a,\frac{1}{2}\,c),\;\; \mathrm{layer\;C}:(\frac{1}{2}\,a,\frac{1}{2\sqrt{3}}\,a,c),\thinspace(0,\frac{2}{\sqrt{3}}\,a,c),\thinspace(a,\frac{2}{\sqrt{3}}\,a,c).$\\
$a$ is nearest neighbor distance in the basal plane and $c/2$ is the distance between two neighbouring layers.}
 \label{fig:fcc_hcp_box}
\end{figure}

Face centered cubic (fcc) and hexagonal close-packed (hcp) are two regular lattices with the highest possible hard--sphere 
packing fraction ($\eta\approx0.74$). The body centred cubic (bcc) structure can attain only packing fractions up to 
$\eta\approx0.68$. 
The fcc and hcp structures differ in how sheets of hexagonally packed hard spheres are stacked upon one another. 
Relative to a reference layer A (see Fig.~\ref{fig:fcc_hcp_box}), two other layer types B and C are possible which are
laterally shifted with respect to A. 
In the fcc structure the stacking of the hexagonally--packed planes corresponds to the crystallographic [111] direction and 
every third layer is the same (ABCABCA) whereas in the hcp lattice ([001] direction), 
the sequence of A and B repeats (ABABABA) (Fig.~\ref{fig:fcc_hcp_box}). If the binding energy (or free energy) 
were dependent only on the number of nearest-neighbor bonds per atom (bonds have no direction), 
there would be no energetic difference between the fcc and hcp structures.

The most convenient unit cell for fcc is the cubic unit cell with $8$ particles at the corners and 6 face-centred particles
(this cell, however, lies oblique in the ABCABCA packing discussed above).
For hcp it is the unit cell with hexagonally packed hard spheres on the basal plane. In order to avoid any numerical errors 
in the comparison between fcc and hcp, we define two extended unit cells of the same size with hexagonally packed spheres as
the base plane (see Fig.~\ref{fig:fcc_hcp_box}). 
Discretizing the extended unit cells by the same number of equal--distant grid-points 
ensures that the lattice points in layer A are on grid points and for layers B and C 
the lattice points are equally "off--grid" since there is a mirror reflection symmetry with respect to the $x$--axis
between B and C. In view of the narrow density peaks centered around each lattice point, this choice eliminates
numerical differences between fcc and hcp free energies to a large extent.
In Fig.~\ref{fig:fcc_hcp_box}, $a$ is the nearest neighbor distance, and in the close-packed case $a=\sigma$. 
The fcc cubic symmetry requires $c=\sqrt{8/3}\;a$ which entails that the distance between nearest neighbors
within a base plane is the same as between neighboring planes. For hcp, the hexagonal symmetry group does not enforce
this constraint, for a discussion of the implications thereof see Sec.~\ref{sec:fcc_hcp} below.

\subsubsection{Free minimization} 

We determine the equilibrium crystal profile $\rho_{\rm eq}(\mathbf{r};\rho_{0},\,n_{\rm vac})$ by a full minimization in 
three--dimensional real space. The density $\rho(\mathbf{r})$ is discretized over a cuboid volume with  
edge lengths $L_{x}=L_{y}=L_{z}=a$ for bcc (using the cubic unit cell) 
and $L_{x}=a,\;L_{y}=\sqrt{3}\,a$ and $L_{z}=3c$ for both fcc and hcp (using the extended unit cells of
Fig.~\ref{fig:fcc_hcp_box}) with 
periodic boundary conditions. We perform a 
double step minimization of the free energy. In the first step, the bulk density $\rho_0$ and 
the vacancy concentration $n_{\rm vac}$ are fixed and the Euler--Lagrange equation (\ref{ele.eq}) is solved iteratively
with a start profile given by the Gaussian profile (\ref{eq:gauss}) with optimal width.
The excess chemical potential $\mu^{\rm ex}$ in Eq.~(\ref{ele.eq}) is treated as a Lagrange multiplier to ensure
the constraint of fixed $n_{\rm vac}$.  
In the next step, this procedure is repeated for different $n_{\rm vac}$ (still keeping $\rho_0$ fixed), 
and the equilibrium density profile 
is determined by minimizing the free 
energy per particle with respect to the the vacancy concentration, $n_{\rm vac}$. For a more detailed discussion of
this procedure see Ref.~\cite{Oet10}. 

In the program, the density profile $\rho$ 
and 11 weighted densities (two scalar densities $n_{3}$, $n_{2}$, three vector densities, 
$\vect{\omega}_{2,i},\;i=x,y,z$, and six tensor densities, $\omega^{t}_{ij}$) need to be discretized on a 
three dimensional grid covering the cuboid boxes. Usually we chose grids for the bcc unit cell 
with $64\times 64\times 64$ points in the $x$, $y$ and $z$ directions, respectively, and $128\times 128\times 384$
points for the fcc and hcp extended unit cells. 
Convolutions in real space are multiplications 
in Fourier space. The necessary convolutions are computed using Fast Fourier Transformations. We use the FFTW 3.3 library 
for parallelized Fast Fourier Transforms. The other parts of the code are parallelized through OpenMP.

There are many sophisticated algorithms for minimizing a function and likewise many techniques to increase 
the speed and efficiency of the process. To have a more efficient algorithm, the iteration 
of Eq.~(\ref{ele.eq}) was done using a 
combination of Picard steps and {DIIS} steps (Discrete  Inversion in 
Iterative Subspace)~\cite{Koval99}. In order to prevent the procedure from diverging during the Picard iterations, 
in each step we mix the new density with the old one,
\bea
\rho_{\rm new}=(1-\alpha)\rho_{\rm old}+\alpha \rho_{\rm new}\;.
\eea
Here, $\alpha$ is a mixing parameter  and it is usually a small number. For the case of bcc, $\alpha$ can be adapted in the
course of the iterations in the range of $\alpha=10^{-5}\dots 10^{-3}$. 
For fcc and hcp, a constant value for $\alpha$ stabilizes the iterations, with values $\alpha = 10^{-5}\dots 10^{-4}$. 
A typical FMT run consisted of an initial Picard sequence with about $30$ steps. Then we alternated between Picard sequence of 7 
steps and a DIIS step (which needs another $n_{DIIS}$ Picard initialization steps), see also Ref.~\cite{Oet12}.

\subsection{Stillinger's expansion in correlated, contiguous particles}
\label{sec:still}

\subsubsection{General outline}

Consider the canonical partition function for $N$ hard spheres:
\bea
  Q(N,V,T) &=& \frac{1}{N!\Lambda^{3N}} \int d\vect r_1 \dots \int d\vect r_n \prod_{i,j\,(i<j)}^N \phi(ij) \;, \\
    & & \phi(ij) = \left\{ \begin{matrix} 0 \qquad (r_{ij} \le \sigma) \\
                                         1 \qquad (r_{ij} > \sigma ) \end{matrix}
                   \right. \;.
\eea
Here, 
$r_{ij}=|\vect r_i-\vect r_j|$ is
the center distance between particles $i$ and $j$. We consider a reference lattice of our choice (fcc, hcp or bcc) with
 $M \ge N$ lattice sites at positions $\vect s_i$ spanning the volume $V$. We associate each particle $i$ with a lattice site
at site $\vect s_i$ and that association divides the $3N$ dimensional configuration space into nonoverlapping
regions $\Omega_{l,p}$. The precise form of this association is discussed in Ref.~\cite{Sti65}, but one may think
of it loosely in terms of each particle $i$ belonging to the Voronoi cell around site $\vect s_i$ of the lattice.
For a chosen subset of $N$ lattice sites $\{\vect s_i\}$ and associated cells, the index $p$ runs over the $N!$ permutations 
of the particles among these cells and this  leads to an identical division of the configuration space,
$\Omega_{l,p_1} \equiv \Omega_{l,p_2}$. The index $l$ runs over the different associations of $N$ particles with
$M>N$ lattice sites and becomes important in the case of finite vacancy concentration.
Thus we obtain for the partition function:
\bea
  Q(N,V,T) &=& \frac{1}{\Lambda^{3N}} \sum_l \int \dots \int_{\Omega_{l,1}} d\vect r_1 \dots  d\vect r_N \prod_{i<j} \phi(ij) \;. 
\eea 
For zero vacancy concentration, this decomposition is akin to the SOC method (as e.g. discussed in Ref.~\cite{Woo97})
where each particle is confined to its Wigner--Seitz cell.
Following Ref.~\cite{Sti65}, one may write $Q$ in terms of configuration integrals $Z_i^l$, $Z_{ij}^l$, \dots which 
describe the correlated motion of one, two, \dots particles in a background matrix of $N-1$, $N-2$, \dots particles fixed at their
associated lattice sites. These configuration integrals are defined as
\bea
  Z_i^l &=& \int_{\omega_i^l} d\vect r_i \prod_{j \not = i}^N \phi(ij) \qquad {\rm with} \; \\
      & & \vect r_j= \vect s_j \quad (j\not =i) \;,  \nonumber \\ 
  Z_{ij}^l &=& \int_{\omega_{ij}^l} d\vect r_i d\vect r_j \prod_{k \not = i,j}^N \phi(ik) \phi(jk) \qquad {\rm with} \; \\
      & & \vect r_k= \vect s_k \quad (k\not =i,j) \;,  \nonumber \\
      & \vdots & \nonumber  \;.
\eea 
The integration domains must fulfill $\omega_{i}^l, \omega_{ij}^l, \dots \in \Omega_{l,1}$, and they depend on the indices of
the free particles $i,j$ and also in the index $l$ determining at which lattice sites the other particles are fixed. 
The partition function is now expressed as the product
\bea
  Q(N,V,T) &=& \frac{1}{\Lambda^{3N}} \sum_l \prod_i^N Z_i^l \; \; \prod_{i<j}^N \frac{Z_{ij}^l}{Z_i^l Z_j^l} \; \;
                        \prod_{i<j<k}^N \frac{ Z_{ijk}^l \, Z_i^l Z_j^l Z_k^l }{Z_{ij}^l Z_{ik}^l Z_{jk}^l}  \dots \\
           & =:& \frac{1}{\Lambda^{3N}} \prod_i^N Y_i^l \; \; \prod_{i<j} Y_{ij}^l  \; \; \prod_{i<j<k}^N  Y_{ijk}^l \dots \;.
\eea
The $Y'$s can also be expressed by the recursive relation
\bea
  Y_{1\dots n}^l &=& \frac{ Z_{1\dots n}^l }{ \prod_{\rm subsets} Y_{i_1 \dots i_m}^l } \;,
\eea
where $\{i_1 \dots i_m\}$ is any proper subset of $\{1\dots n\}$. (For example, when omitting indices we have
$Y_2 = Z_2/(Y_1Y_2)$ and $Y_3 = Z_3/(Y_1Y_2Y_3\,Y_{12}Y_{13}Y_{23})$.)  

\subsubsection{Expansion up to $n=2$ for hcp, fcc and bcc hard spheres }

In the following, we restrict calculations to the case $N=M$ (number of particles equal to number of lattice sites), 
i.e. consider a vacancy--free crystal. From simulations \cite{Ben71} and FMT \cite{Oet10} we can estimate that the effect of 
vacancies on the free energy of the crystal is small: for fcc hard spheres we have $n_{\rm vac} \sim 10^{-4}$
(simulations) and $n_{\rm vac} \sim 10^{-5}$ (FMT) in equilibrium at coexistence, the corresponding free energy shift 
compared to $n_{\rm vac} \to 0$ can be estimated from FMT, $\Delta F/N \sim 10^{-5}$ $\kt$.   

Truncated after the first term, the Stillinger series  is
\bea
   Q_1 &=& \frac{1}{\Lambda^{3N}} (V_1)^N\;,
\eea 
where $Z_1^l$ has been reduced to $V_1$, the free volume for one particle in a cage of fixed neighbors at
their lattice sites. Consequently the free energy is
\bea
  \beta F_1 &=& - N \ln \frac{V_1}{\Lambda^3}\;.
\eea
For fcc and hcp, $V_1$ is equal and has been calculated analytically in Ref.~\cite{Bueh51}, we quote this result in App.~\ref{app:v1}. 
For bcc, we did not find a literature
result and therefore give the calculation and result also in App.~\ref{app:v1}. 

The second term in the Stillinger series for $Q$ gives only a contribution different from 1 if the
two fixed particles are neighbors. Thus the truncated Stillinger series is
\bea
   Q_2 &=& \frac{1}{\Lambda^{3N}} (V_1)^N \prod_k \left( \frac{V_{2,k}}{(V_1)^2} \right)^{g_k N} 
\eea 
Here, $V_{2,k}$ is the correlated free volume of the two neighboring particles (with dimension (length)$^6$) 
which may depend on the type of neighbor configuration (index $k$). The power $g_k N$ reflects the freedom to choose the first
of the two particles to be any of the $N$ particles in the system and $g_k$ is the multiplicity of the 
neighbor configuration. It is half the number of neighbors of type $k$ for a given fixed particle. 
The associated free energy is 
\bea
   \beta F_2 &=& \beta F_1 - N  \sum_k g_k \ln \left( \frac{V_{2,k}}{(V_1)^2} \right).   
\eea
For our considered
lattice cases the neighbor types and multiplicities are given in Tab.~\ref{tab:neighbor}. The cubic lattices fcc and
bcc have only one neighbor type whereas for hcp there is a difference whether the neighbor is within the same
close--packed plane or in an adjacent close--packed plane. See also Ref.~\cite{Sti68} for the multiplicities 
corresponding to the third term in the series (fcc and hcp).    

\begin{table}
 \begin{center}
  \begin{tabular}{llll} \hline \hline
    lattice$\qquad$  & neighbor type & $k$ $\qquad$& $g_k$ $\qquad$\\ \\ \hline   
     fcc     &   all neighbors & 1 & 6 \\
     hcp     & within close-packed plane & 1 & 3 \\
             & in adjacent close--packed planes & 2 & 3 \\
      bcc    & all neighbors & 1 & 4 \\ \hline \hline
  \end{tabular}
 \end{center} 
 \caption{Neighbor configurations with multiplicities for the different lattices.}
 \label{tab:neighbor}
\end{table}

We calculate the two--particle volumes $V_{2,k}$ for different densities 
by a simple Monte--Carlo computation. For that we specify a suitably large cuboid volume $V_c$ for each of the
two free particles from which $n$ sets of random positions (for each of the two particles) are drawn. 
For each set of random positions overlap is checked
with the other particle and the fixed neighboring particle, leading to a total of $n'$ sets of random positions with
no overlap. Then $V_{2,k} = (n'/n) V_c^2$. The statistical error  $\Delta V_{2,k}/V_{2,k}$ needs to be below
$10^{-5}$ for a reliable assessment of the free energy difference between fcc and hcp, and this is achieved 
with 1000 subsets, each containing $n=10^9$ sets of random positions. In the limit $\rho_0 \to \rho_{\rm cp}$
($\rho_{\rm cp}=\sqrt{2}/\sigma^3$ is the close--packing density) agreement was found with the analytical results
of Ref.~\cite{Sti68}, but we had to approach $\rho_{\rm cp}$ very closely to establish that.

\section{Results}

\label{sec:results}

\subsection{Stillinger series}

\begin{figure}
  \centerline{\epsfig{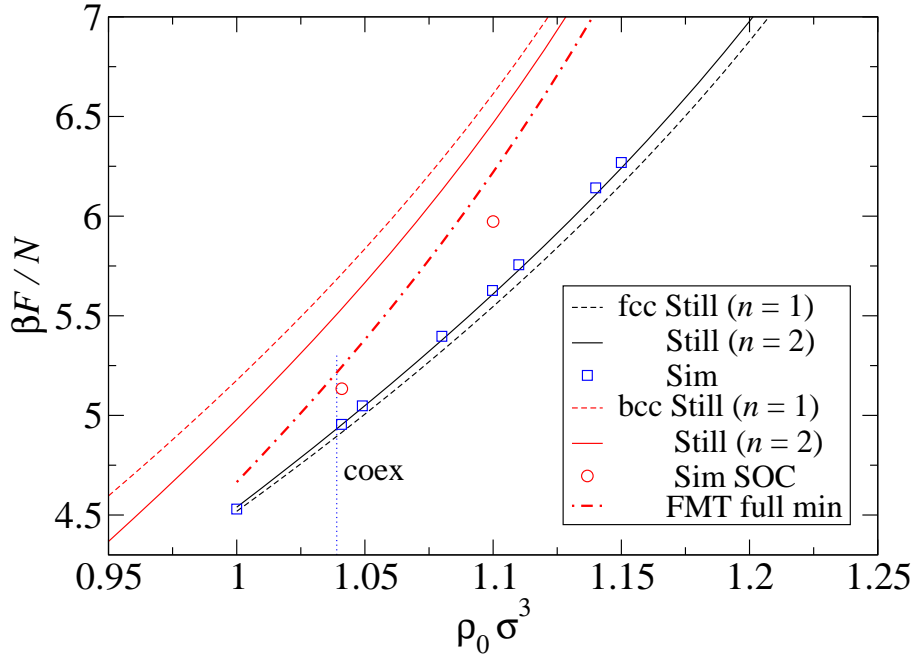} }
  \caption{(color online) Crystal free energies $\beta F/N$ for fcc and bcc from the Stillinger series in comparison to simulation data and FMT results (bcc).
      For fcc, simulation data are taken from Refs.~\cite{VegaNoya07,Oet10}, and for bcc, simulation data are obtained using
      the single--occupancy cell method (SOC) \cite{Run87}. The FMT data are this work, see Sec.~\ref{sec:bcc}.  }
  \label{fig:f_still}
\end{figure}

For fcc and hcp, the Stillinger series truncated at $n=2$ gives very good results for the free energy per particle
$F/N$
(see Fig.~\ref{fig:f_still}, to obtain numbers, we put $\Lambda=\sigma$). We have compared to very precise simulation
data obtained in Refs. \cite{VegaNoya07,Oet10} which have an error of about $0.002$ $\kt$. The Stillinger series ($n=2$) results
for $F/N$ deviate from these ranging from 0.01 $\kt$ (at $\rho_0\sigma^3=1.0$) to 0.03 $\kt$ (at  $\rho_0\sigma^3=1.15$),
this is less than 0.5\% relative deviation. This is about the same accuracy we obtain with FMT (see also Ref.~\cite{Oet10}).
Note, however, that a deviation of the order of 0.01 $\kt$ is about 10 times higher than the fcc--hcp free energy
difference obtained from simulations, as discussed before. 

For bcc, the situation is very much different. Since the bcc structure for hard sphere is unstable against shear, the 
crystal can be stabilized in simulations only by constraints such as in the SOC method.
We would expect from the previous derivation that the Stillinger expansion
is a reasonable series expansion for the free energy of the SOC method. However, as Fig.~\ref{fig:f_still} demonstrates,
the first two terms are quite far away from the SOC data and also from the FMT results for the branch with lowest free energy, 
pointing to the importance of 
higher correlations. (Ultimately, the shear instability is a collective many--body effect, so perhaps the importance of
many--particle correlations also in the constrained crystal is not too surprising.) See, however, the next subsection for 
a more detailed discussion on bcc solutions within FMT, especially with regard to a solution branch with higher free energy 
which appears to be linked to the bcc Stillinger solution. 

Finally, for fcc/hcp the inclusion of the correlated neighbor term {\em increases} the free energy, whereas for bcc
it leads to a  {\em decrease}. 

\subsection{bcc -- FMT results}

\label{sec:bcc}

\begin{figure}
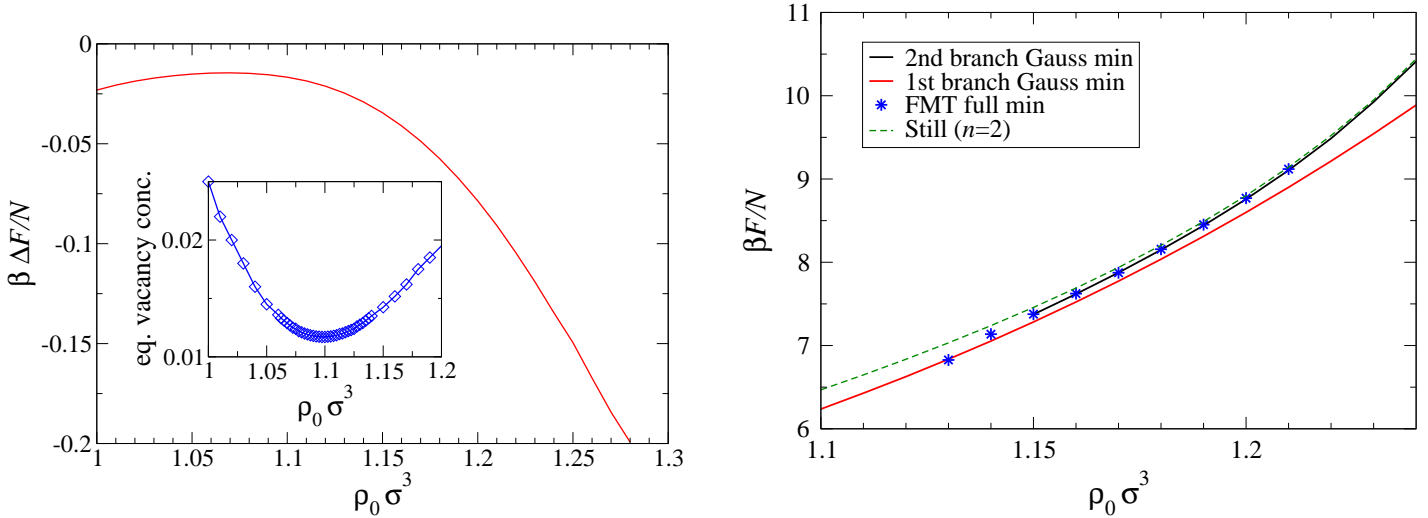

  \centerline{\epsfig{file=bcc_nvac_rho.eps, width=9cm} \hspace{0.5cm}
              \epsfig{file=bcc_2branches.eps, width=9cm} }
  \caption{(color online) (a) Difference in free energy per particle between the fully minimized and the Gaussian solution for the
 first branch of the bcc solutions as a function of bulk density. Inset: Equilibrium vacancy concentration
 as a function of bulk density for the same first branch. (b) Free energy per particle as a function of bulk
 density for the bcc solution of the second branch: Full minimization (symbols, $n_{vac}=6\times10^{-4}$ fixed) and Gaussian approximation (full black line).
 For comparison the Stillinger result ($n=2$) is given (dashed line) as well as the Gaussian approximation for
 the first branch (dot--dashed line). }
  \label{fig:bcc}
\end{figure}

As already discussed, a bcc crystal solution can only be stabilized by constraints. In FMT, these are the periodic
boundary condition on the cubic unit cell.
Within the Gaussian parametrization (see Eq.~(\ref{eq:gauss})), bcc solutions in FMT 
(Rosenfeld, Tensor and White Bear Tensor, see Sec.~\ref{sec:fmt}) 
have been investigated by Lutsko \cite{Lut06}
(with the additional constraint $n_{\rm vac}=0$, such that in the free energy minimization, 
the width parameter $\alpha$ is the only variable which is
varied at a given bulk density $\rho_0$). 
For small bulk densities ($\rho_0\sigma^3 \alt 1.16$), Lutsko found a single free energy minimum with a rather
small width parameter $\alpha\approx 30\dots 40$, indicating a broad Gaussian peak. Interestingly, $\alpha(\rho_0)$ exhibits
a maximum at $\rho_0\sigma^3 \approx 1.13$ and then {\em decreases} again upon increasing the density (i.e.
the density peaks become wider upon compressing the crystal!). Moreover, at bulk densities $\rho_0\sigma^3 \agt 1.16$ a second free 
energy minimum was visible (with higer free energy). In this second branch, the width parameter increased (the peak width decreased) 
with increasing density as one would naively expect.  

We investigate these findings further using full minimization. For the first branch with lowest free energy, we confirm
that there is a minimal width of the peaks at $\rho_0\sigma^3 \approx 1.13$. Full minimization reveals a rather strong 
deviation from the simple Gaussian form 
in the density peaks: The difference in free energy per particle $F/N$  between Gaussian and full minimization is about 0.1 $\kt$
(see Fig.~\ref{fig:bcc} (a)) and thus
about 2 orders of magnitude higher than in the case of fcc \cite{Oet10}. Curiously, this free energy difference
increases with increasing density beyond $ \rho_0\sigma^3 \approx 1.07$. Secondly, the equilibrium vacancy concentration
$n_{\rm vac}$ is of the order of 10$^{-2}$ and thus {\em several orders of magnitude higher} than found in fcc.
$n_{\rm vac}(\rho_0)$ has a minimum at $ \rho_0\sigma^3 \approx 1.10$ and then increases again, adding to the peculiarities
of this solution branch. We note that in an FMT study of parallel hard squares and cubes similar peculiarities have been found
\cite{Bel11}.    

The second branch found by Lutsko is not an artefact of the constrained Gauss minimization. By a careful iteration procedure,
we found corresponding fully minimized solutions whose free energy per particle is very close to the values from the
Gaussian approximation (thus very much like the fcc solutions and very much unlike the solutions from the first branch),
see Fig.~\ref{fig:bcc} (b). For increasing densities, we see a convergence of $F/N$ to the results
of the Stillinger series ($n=2$). Thus the second branch of the bcc solutions has the same character as the fcc solution when 
compared with the Stillinger approach: only a few correlated particles are sufficient to obtain the free energy.

One could argue that the discussion of these bcc solutions is futile and void of physical significance  in view of their  
overall instability. However, the quality of the FMT functionals and their success in describing the fcc phase
leads us to think that these solutions are perhaps not to be discarded altogether. Since around coexistence 
($\rho_0\sigma^3 \approx 1.04$) the difference in $F/N$ to the fcc crystal is about 0.3 $\kt$ and thus very high, it is
reasonable that bcc crystallites have not been observed in the nucleation process of a hard sphere crystal. Nevertheless,
the bcc solutions are perhaps a useful reference point for discussing the crossover from fcc to bcc as the most stable crystal
structure
for other potentials such as of $(\sigma/r)^n$ type. These could be treated by suitable perturbation ansatz in the free energy
functional. Also, it could be interesting to investigate further the dispersion relation of phonons for the solutions of the
first branch and thus shed further light on the shear instability. 

\subsection{fcc/hcp: Free energy differences and density anisotropies}

\label{sec:fcc_hcp}

\begin{figure}
  \centerline{\epsfig{file=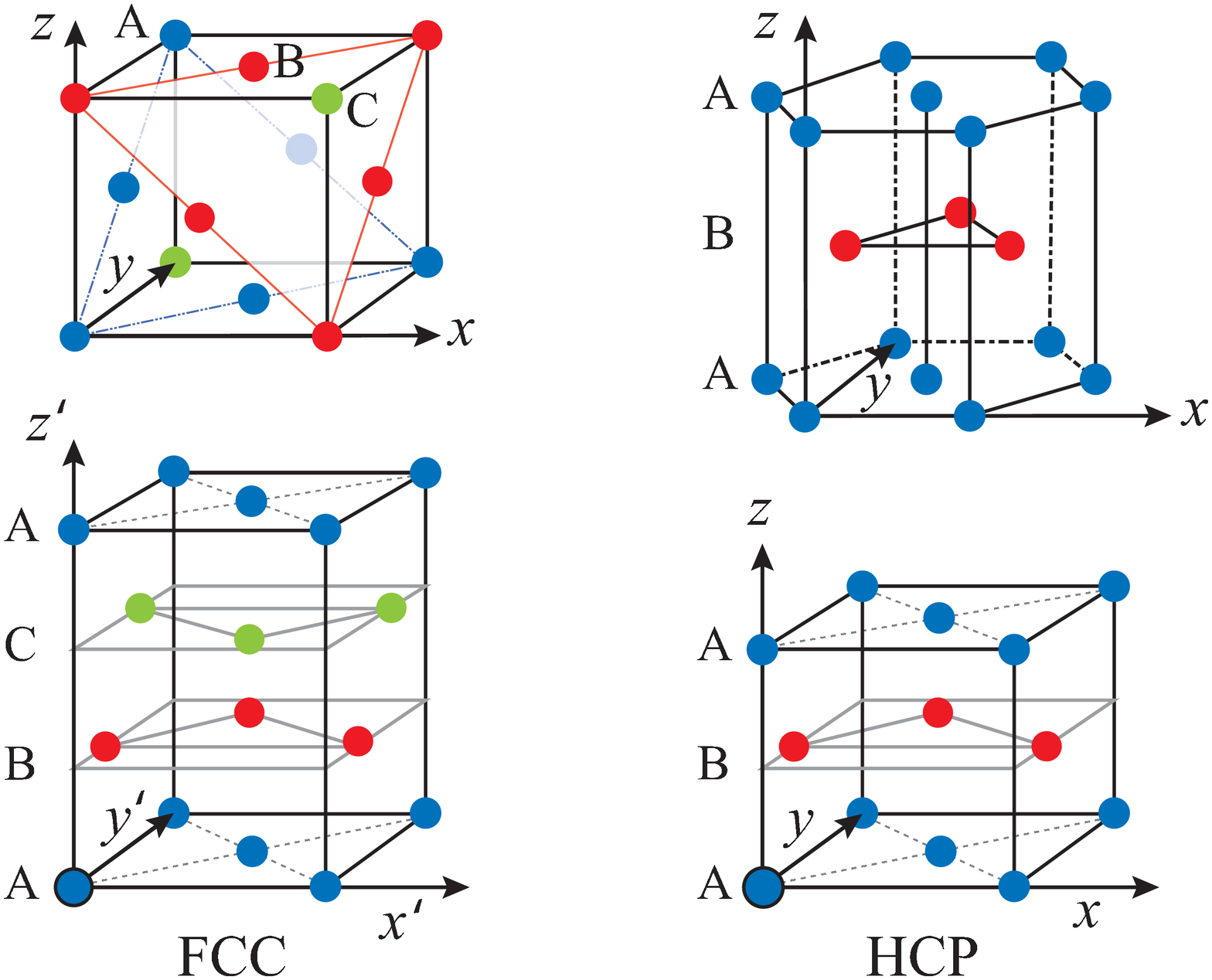, width=7.2cm} \hspace{0.5cm}
              \epsfig{file=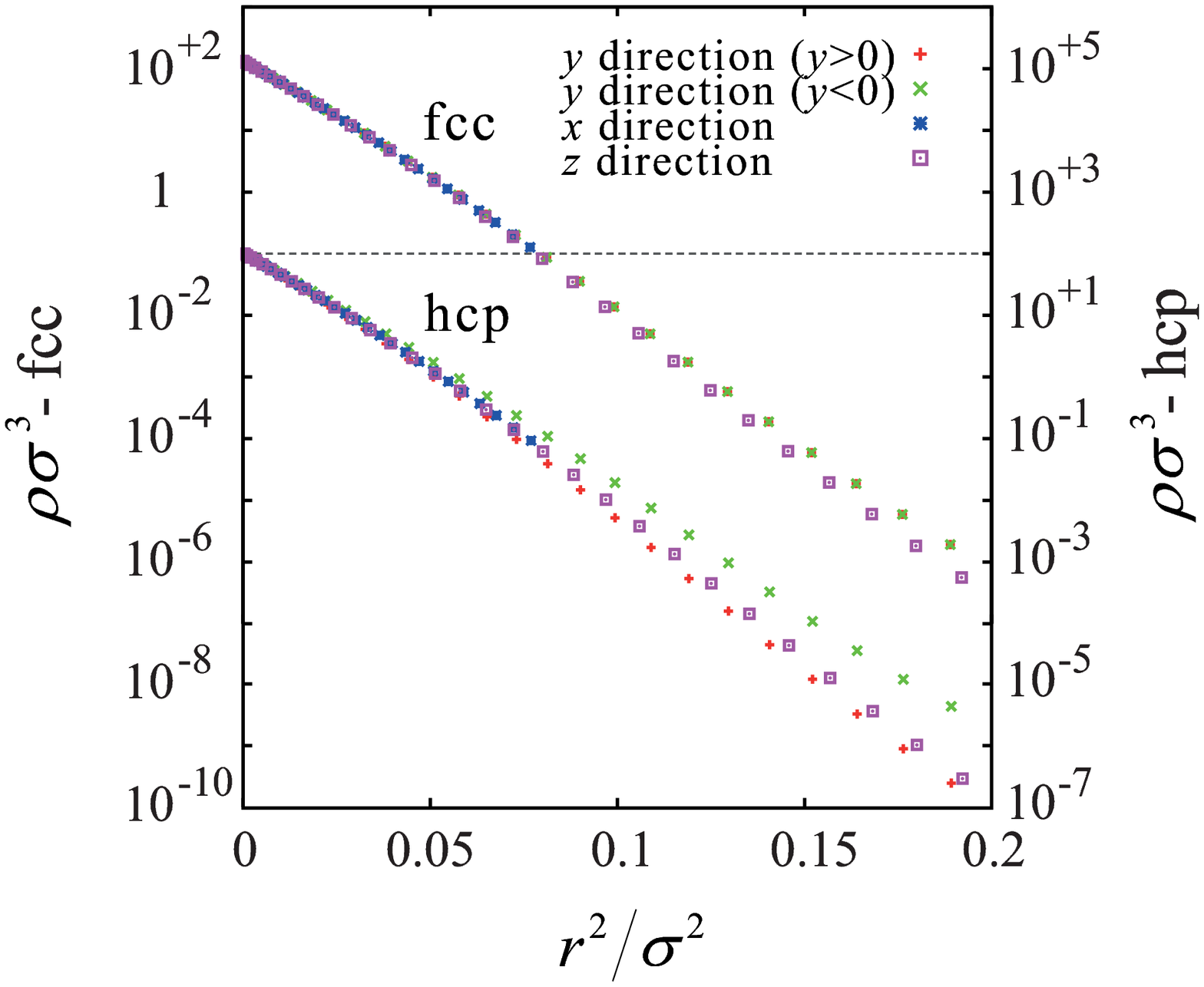, width=7cm} \hspace{0.5cm} }
  \caption{(color online)  Unit cells and density anisotropies for fcc and hcp. (a1) and (a2) show the most convenient
 unit cells (cubic for fcc and hexagonal for hcp) for the mathematical discussion of the density anisotropies
 (see Eqs.~(\ref{eq:fcc_aniso}) and (\ref{eq:hcp_aniso})). (a3) and (a4) show the unit cells used in the
  numerical computations. The hexagonally packed planes (marked in different colors) lie oblique in the cubic unit cell (a1).
 (b) fcc and hcp density distributions around the  lattice site at the origin in different directions.
  Here, we used the bulk density $\rho_0\sigma^3=1.04$ and fixed the vacancy concentration to $n_{\rm vac}=10^{-4}$. }
  \label{fig:aniso}
\end{figure}

As discussed in the Introduction, FMT gives the same free energy per particle $F/N$ for fcc and hcp when the
Gaussian approximation is employed \cite{Tar00,Lut06}. Free minimization lifts this degeneracy in the free energy.
In order to understand this result qualitatively, it is useful to consider the symmetries in the unit cell
of fcc/hcp and the constraints these symmetries place upon the lattice--site density profiles. For fcc, this is best 
discussed by considering the cubic unit cell in Fig.~\ref{fig:aniso} (a1). The non--radial contributions to the 
density profile  around the lattice
point in the origin can be expanded in a Taylor series in $x,y, z$ where the terms in this series must respect
the 48 point symmetry operations in the cubic unit cell (belonging to point group
$\frac{4}{m}\bar 3 \frac{3}{m}$ in Hermann--Mauguin notation) \cite{Oet10}:
\bea
  \rho_{\rm fcc} (x,y,z) = \rho_{\rm rad}(r)  \; (1 + K_4(x^4 + y^4 + z^4) + \dots ) \;.
  \label{eq:fcc_aniso}
\eea  
Here, $\rho_{\rm rad}(r)$ is an averaged, radial profile which is more or less of Gaussian shape. 
The leading anisotropic term is of polynomial order 4 with expansion coefficient $K_4$. One can also understand this result by resorting to an expansion
in the subset of spherical harmonics which respect the cubic point symmetry, 
this leads to an expansion in the so--called Kubic Harmonics \cite{Bet47}. -- 
For hcp, we consider the unit cell in Fig.~\ref{fig:aniso} (a2). The corresponding Taylor expansion for the
non--radial contributions to the density profile  around the lattice
point in the origin has to respect only the 24 point symmetry operations appropriate for 
the hexagonal group $\frac{6}{m}\frac{2}{m}\frac{2}{m}$. According to Ref.~\cite{Niz76}, this leads to 
\bea
  \rho_{\rm hcp} (x,y,z) = \rho_{\rm rad}(r) \; (1 + K'_2 z^2 + K'_3 y (3x^2 - y^2) + \dots)\;,
  \label{eq:hcp_aniso}
\eea 
where polynomial terms up to order 3 have been taken into account (with expansion coefficients $K'_i$).
The corresponding construction using spherical harmonics leads to the so--called Hexagonal Harmonics.
We observe that there is a qualitative difference in the shape of the density profile between hcp and fcc
according to these expansions:
\begin{itemize}
 \item[$(i)$:] To leading order in anisotropy for hcp, the density peak $\rho(r)$ should look different
    in $z$--direction (perpendicular to the  hexagonally packed planes) than in directions in the $x$--$y$ 
    plane. To phrase it differently: one would expect different width parameters $\alpha_z, \alpha_{x,y}$ for 
    a Gaussian density peak of the form $\rho_{\rm hcp} (x,y,z)  \propto \exp( - \alpha_{x,y}(x^2+y^2) - \alpha_z z^2)$. 
    We did not observe
    this in our numerical solutions but we will return to this point below. 
 \item[$(ii)$:] To next--to--leading order in the anisotropy for hcp, we expect a different behavior when
    comparing $\rho(0,y,0)$ with $\rho(0,-y,0)$ due to the antisymmetric term $\propto K'_3$ in Eq.~(\ref{eq:hcp_aniso}).
    Such a symmetry breaking is not present in the fcc peak. To demonstrate this difference, we compare 
    $\rho(0,\pm y,0)$, $\rho(x,0,0)$, and $\rho(z,0,0)$ between fcc and hcp, see Fig.~\ref{fig:aniso} (b) and (c).\footnote{
    Note that in our numerical computations we used 
    the unit cells depicted in Fig.~\ref{fig:aniso} (a3) (fcc),
    and in Fig.~\ref{fig:aniso} (a4) (hcp). Thus, the fcc cubic unit cell and the unit cell in Fig.~\ref{fig:aniso} (a3)
    are related by a three--dimensional rotation. Likewise, the anisotropy expansion for the extended unit cell must be obtained
    from the corresponding expression  (\ref{eq:fcc_aniso}) for the cubic unit cell by applying this rotation. However,
    since the density anisotropy is $\propto y^4$ ($x=0,z=0$) in Eq.~(\ref{eq:fcc_aniso}), the corresponding density anisotropy
    must also be $\propto y'^4$ ($x'=0,z'=0$) in the rotated unit cell 
    (primes denote the coordinates in the extended unit cell in  Fig.~\ref{fig:aniso} (a3)).} 
    Indeed we observe that the symmetry is broken for the hcp profile, in accordance with the anisotropy expansion, and we conclude
    that the fcc/hcp free energy difference in FMT results from this symmetry breaking.
\end{itemize}

\begin{figure}
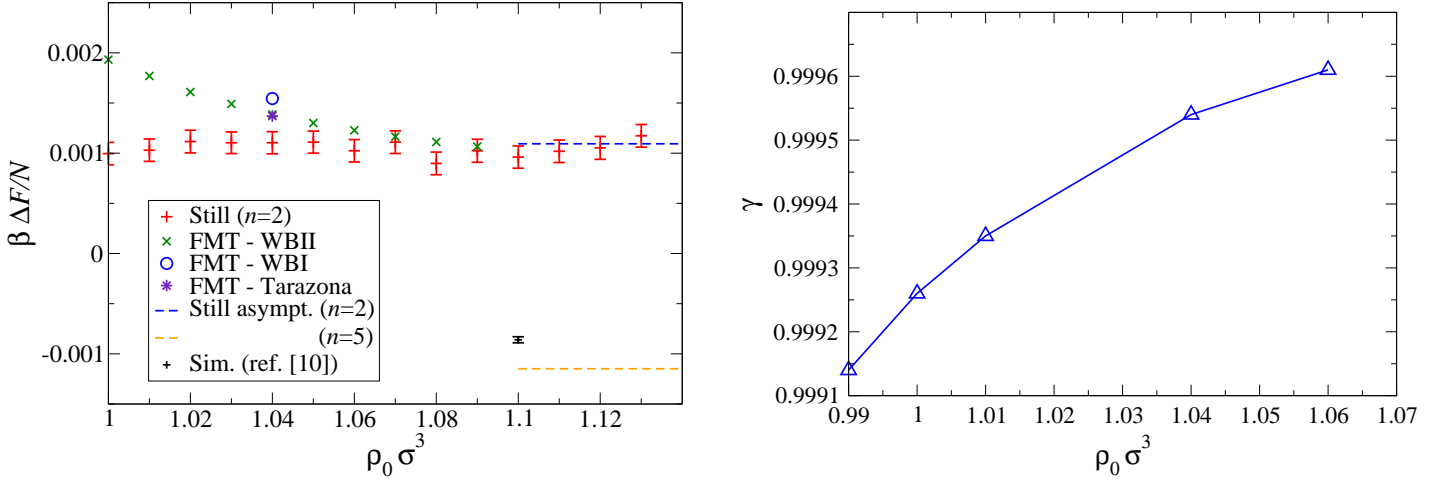

  \centerline{\epsfig{file=master_plot.eps, width=9cm} \hspace{0.5cm}
              \epsfig{file=fcc_cell_resize.eps, width=9cm} }
 \caption{(color online) (a) Free energy difference between fcc and hcp vs. bulk density. The black symbol shows the simulation value from Ref.~\cite{Wil97}. Rest of the symbols show the data obtained from FMT and 
 the Stillinger series ($n=2$) and dashed 
 lines show the asymptotic behavior of the free energy difference near close packing for the Stillinger series (different $n$) 
 \cite{Koch05}. (b) Distortion parameter $\gamma=c/c_0$ which mimimizes the hcp free energy
 vs. bulk density. In all FMT calculations we put $n_{\rm vac} = 10^{-4}$. }
  \label{fig:hcp_free_energies}
\end{figure}

Our results for the fcc/hcp free energy difference per particle are given in Fig.~\ref{fig:hcp_free_energies}(a). In FMT (White Bear II--Tensor), the
difference $\beta \Delta F/N$ is larger than zero, implying that hcp has lower free energy. Furthermore, there is only a moderate
drop of $\beta \Delta F/N$ with the bulk density $\rho_0$. At coexistence ($\rho_0\sigma^3=1.04$), we have computed $\beta \Delta F/N$
also for other FMT functionals (Tarazona--Tensor, White Bear--Tensor) and found no change in sign but a variation in magnitude by 50\% or
$5\cdot 10^{-4}$. In view of the variation of $\beta F/N$ for fcc between the functionals (about  $4 \cdot 10^{-2}$, i.e. a factor of 80 larger),
the functionals are very consistent with each other with respect to the stability of hcp. The results from the Stillinger series ($n=2$) 
for $\beta \Delta F/N$ are approximately constant ($\sim 1 \cdot 10^{-3}$) with increasing density and coincide with the analytical value
at close packing obtained in Ref.~\cite{Sti68}. It is remarkable that also the FMT results seem to converge to this value. --
For comparison, in Fig.~\ref{fig:hcp_free_energies}(a) we have also included the analytical value from the Stillinger series 
($n=5$) \cite{Koch05} and the simulation value of Ref.~\cite{Wil97}. Although FMT does not agree with the sign of $\beta \Delta F/N$ obtained
in the simulation, it is gratifying to note that according to these results FMT is correct on the level of two correlated particles in the
Stillinger picture.   

Finally, we return to the observation that in the hcp density anisotropy the leading term $\propto z^2$ (see Eq.~(\ref{eq:hcp_aniso}))
was missing in our numerical solutions. This is related to our choice of the distance between the hexagonally packed layers
($c/2 = c_0/2=\sqrt{2/3}a$ where $a$ is the nearest neighbor distance, see Fig.~\ref{fig:fcc_hcp_box}). With this choice the distance between
nearest neighbors is the same for two sites within the same  hexagonally packed planes and 
two sites in two adjacent planes. However, the hcp symmetry group 
does not require this, and one is free to choose another distance between the planes. With a different choice, also the
nearest neighbor distance is different for sites in two different planes and also the width of the lattice site
density profiles will be different in the direction normal to the hexagonally packed planes.    
We have investigated whether also the free energy minimum for hcp shifts to a value different from $c_0$. In order to keep
the bulk density constant we defined a stretching parameter, $\gamma=c/c_0$, which describes the distortion of the crystal
in $z$--direction. In order to keep the bulk density constant, we rescaled the nearest neighbor distance in the planes
as follows: $a' = a/\sqrt{\gamma}$. Full minimization was done for a range of $\gamma$ values. The result
for $\gamma$ which minimizes $F/N$ is shown in Fig.~\ref{fig:hcp_free_energies} and it is seen that the equilibrium distortion is
quite small, below $10^{-3}$. The corresponding free energy shift per particle compared to the solution with $c=c_0$ is about
$10^{-5}$ $\kt$. These results are actually similar to the ones in Ref.~\cite{Sti01}: There, a similar lattice distortion
was calculated for the zero--temperature Lennard--Jones hcp crystal by lattice sums.

\section{Summary and conclusions}

\label{sec:summary}

In this work we have performed a study of bcc, fcc and hcp hard sphere crystals using
unrestricted minimization in 
density functional theory (DFT) of Fundamental Measure type (FMT) which is currently the most accurate
approach. We have complemented these investigations with an approach which is based on the expanding 
the crystal partition function in terms of number $n$ of free particles while the remaining particles 
are frozen at their ideal lattice positions (Stillinger series). 

For the metastable bcc crystal, we have found two solutions for bcc crystals whose free energies are well above 
the free energies of fcc/hcp (see Fig.~\ref{fig:f_still} and \ref{fig:bcc}(b)). 
The first solution (with a rather large density peak width at lattice sites) 
is characterized by a rather large equilibrium vacancy concentration ($\sim 0.01$) and its free energy
can not be described by the Stillinger approach. The shear instability of bcc is presumably related  to this 
first solution. The second solution (characterized by a small peak width and small equilibrium vacancy concentrations) 
agrees well with the solution from the Stillinger approach ($n=2$) with respect to its free energy.

The free energy degeneracy between fcc and hcp, found in previous approaches using constrained, rotationally--symmetric
density peaks around lattice sites, is broken upon full minimization. The density asymmetries are qualitatively different
for fcc and hcp and agree with expansions in respective lattice harmonics (see Fig.~\ref{fig:aniso}).  
We found that in
FMT the free energy per particle is lower for hcp than the one for fcc by about $10^{-3}$ $\kt$.
This agrees remarkably well with the  Stillinger solution for $n=2$ (see Fig.~\ref{fig:hcp_free_energies}).
Simulations, however, indicate that fcc has a lower free energy than hcp by about the same figure. 
Previous investigations of the Stillinger approach in the high--density limit (near close packing) have shown
that hcp is more stable than fcc for $n=2\dots 4$ and the situation reverses for $n=5$.
Thus, the stability of fcc seems to be a subtle effect involving the correlated motion of at least 5 particles which
currently can not be captured by the FMT functionals.

\begin{appendix}

\section{One--particle volumes for the fcc/hcp and bcc hard--sphere crystal}
\label{app:v1}

\subsection{fcc and hcp}

The one--particle free volume is equal for fcc and hcp and has been given in Ref.~\cite{Bueh51}. 
We introduce the nearest neighbor distance $d = 2^{2/3}\rho_0^{-1/3}$. The hard sphere diameter is $\sigma$ and
the formula is valid for densities $ \rho_0\sigma^3 \in [1/2,  \sqrt{2}]:$
\bea 
 V_1 &=& \frac{20}{3} c^3 - \frac{4}{3}c^2 s - 4c^2\sqrt{\sigma^2-c^2} +  \nonumber \\
     & &    2\sqrt{2} (c^3-6c \sigma^2)\left(\arcsin\frac{c}{q} + \arcsin m\right) + \\
     & &    8\sigma^3 \left( 2\arcsin u + \frac{\pi}{2} - \arcsin w - \arcsin t   \right) \;. \nonumber \\
   \mbox{with} \nonumber \\
   & & c = d/\sqrt{2}\;,  \nonumber \\
   & & s = \sqrt{3\sigma^2-2c^2} \;,  \nonumber \\
   & & q = \sqrt{2\sigma^2-c^2} \;. \nonumber \\
   & & m = (c-2s)/(3q)\;,  \nonumber \\
   & & t = (\sigma^2+c\sigma-c^2)/(q\sigma)\;,  \nonumber \\
   & & u = [(2\sigma+c)(\sigma+[2c-s]/3) - (\sigma+c)^2]/[q(\sigma+[2c-s]/3)] \;,  \nonumber \\
   & & w = (\sigma^2- c\sigma - c^2)/(q\sigma)\;,  \nonumber 
\eea
The shape of the free volumes is sketched in Fig.~\ref{fig:v1}. 

\subsection{bcc}

\begin{figure}
 \epsfig{file=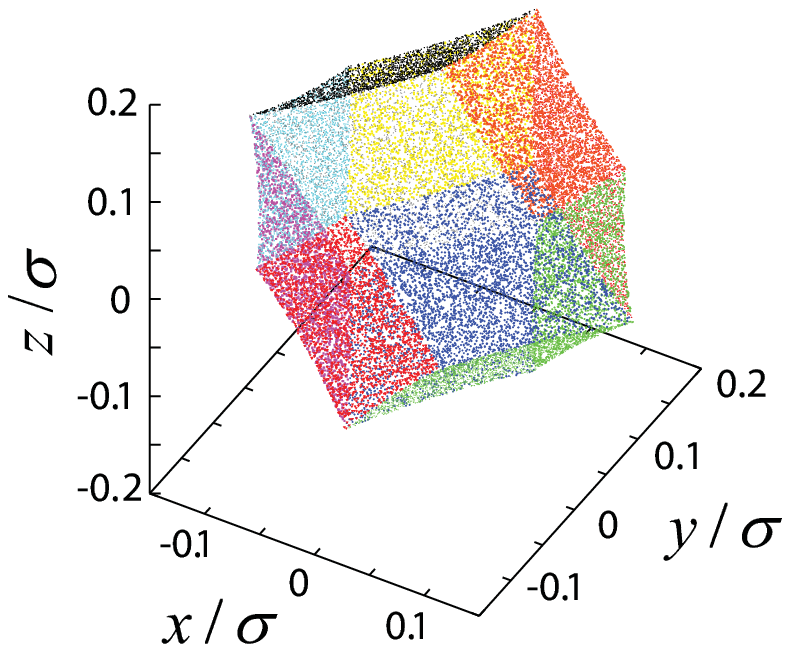, width=5cm} \hspace{0cm}
 \epsfig{file=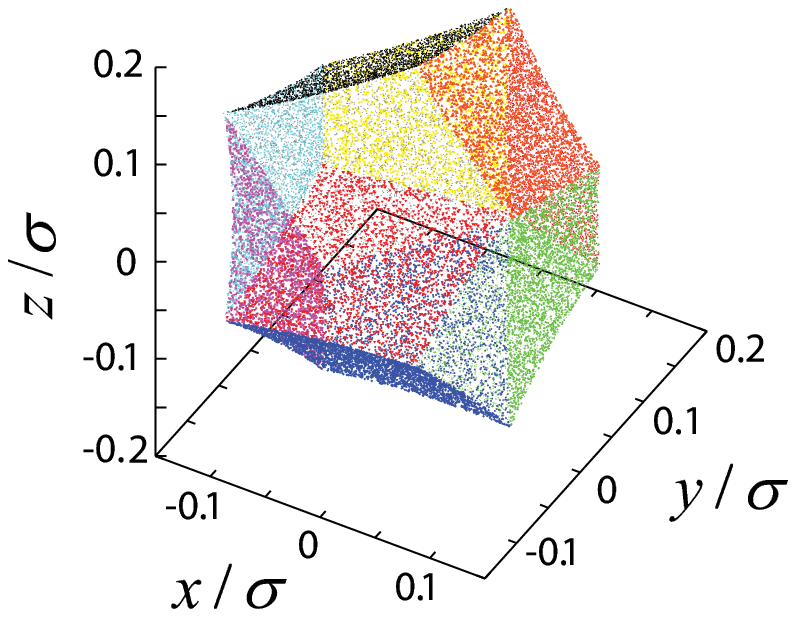, width=5cm} \hspace{0cm}
 \epsfig{file=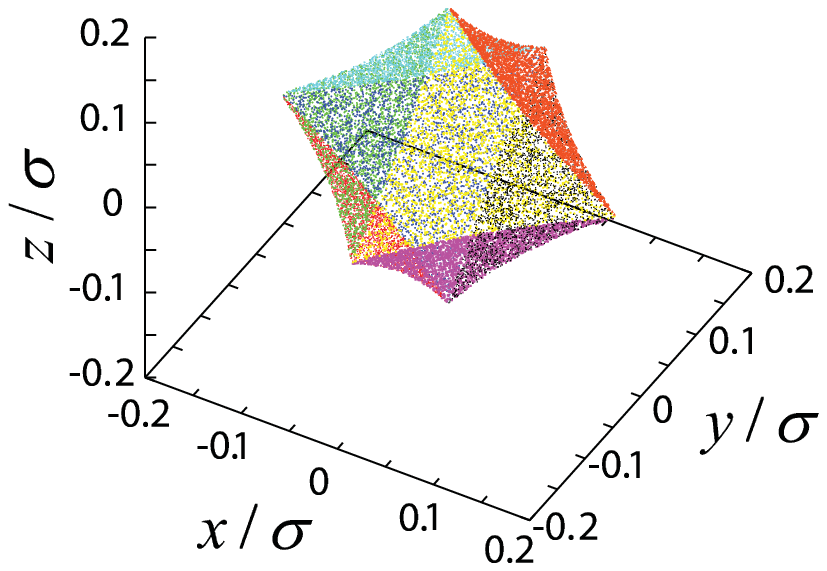, width=5cm} 
 \caption{(color online) Shape of one--particle free volumes for fcc, hcp and bcc (from left to right) at a crystal density of $\rho_0\sigma^3=1$. }
 \label{fig:v1}
\end{figure}

In case of bcc the free volume is given by an octahedral--like body (see Fig.~\ref{fig:v1}) centered in the cubic unit cell.
The faces are parts of the surfaces of the exclusion spheres (of radius $\sigma$) around the corners of the cubic unit cell.
Let $a = (2/\rho_0)^{1/3}$ be the side length of the cubic unit cell. The free volume is then given by
\bea
  V_1 &= & 8 \int_0^{z_{\rm max}} dz \int_0^{x_{\rm max}} dx  \left( \frac{a}{2} - \sqrt{\sigma^2 - \left(\frac{a}{2}-z\right)^2 - \left(\frac{a}{2}-x\right)^2 }\right)  \;, \\
     & & \quad x_{\rm max} =  \frac{a}{2} - \sqrt{\sigma^2 - \left(\frac{a}{2}-z\right)^2 - \frac{a^2}{4} } \nonumber \;, \\
     & & \quad z_{\rm max} =  \frac{a}{2} - \sqrt{\sigma^2 - \frac{a^2}{2} } \nonumber \;. \\
     &=& \frac{a^3}{8} +a \left(\frac{3}{2}\sigma^2-\frac{1}{8}a^2\right) \left( \arctan\frac{2c}{a} - \frac{\pi}{4} \right) - \nonumber \\
    &&  \frac{a^2}{4} c + \frac{2}{3}\sigma^3 \left( \arctan\frac{a^2}{4\sigma c} -  \arctan\frac{c}{\sigma}    \right)\;, \\
     & & \quad c = \sqrt{\sigma^2-a^2/2} \nonumber \;.
\eea

\end{appendix}

\end{document}